\documentstyle[12pt,epsfig,floatfig]{article}
\def\ZO{Z$^0$}
\def\ro{$\rho^0$}
\def\ph{$\phi$}
\def\fo{$f_0$(980)}
\def\fd{$f_2$(1270)}
\def\fdp{$f^{\,'}_2$(1525)}
\def\KOV{K$^{*0}$(892)}
\def\KOT{K$^{*0}_2$(1430)}
\def\xp{$x_p$}
\begin{document}
\initfloatingfigs
{~~~~}

\bigskip
\bigskip
\bigskip
\smallskip

{\large 
Production properties of the orbitally excited mesons \fo, \fd, \KOT\ and 
\fdp\ in \ZO\ hadronic decays}  

\smallskip
\smallskip
\smallskip
{Vladimir UVAROV}

\smallskip
\smallskip
\smallskip
{Institute for High Energy Physics (IHEP), Protvino, Moscow region,
Russia\\} 

\smallskip
{\quad
DELPHI results are presented on the inclusive production of the neutral mesons
\ro, \fo, \fd, \KOT\ and \fdp\ in hadronic \ZO\ decays. They are based on 
about 2 million multihadronic events collected in 1994 and 1995, using the 
particle identification capabilities of the DELPHI Ring Imaging Cherenkov 
detectors and measured ionization losses in the Time Projection Chamber. The 
total production rates per hadronic Z$^0$ decay have been determined to be:
$1.19 \pm 0.10$ for \ro ;
$0.164 \pm 0.021$ for \fo ;
$0.214 \pm 0.038$ for \fd ;
$0.073 \pm 0.023$ for \KOT ;
and $0.012 \pm 0.006$ for \fdp . 
The total production rates for all mesons and differential cross-sections 
for the \ro, \fo\ and \fd\ are compared with the results of other LEP 
experiments and with models.\\}

\smallskip
{\bf 1. INTRODUCTION}

\smallskip
\smallskip
\smallskip

\quad
At LEP it has been observed that a large fraction of primary produced 
particles in hadronic \ZO\ decays are the orbitally excited states ($L$=1).
For certain particles a significant disagreement with currently existing models 
has been observed. 
This is due to the fact that the physics of hadronization is not completely 
understood and these models cannot give sufficient guidance on possible 
differences in the production mechanisms of different mesons and on their 
dependences on spin and orbital momentum dynamics. 
In this view, studies of the production properties of the orbitally excited 
states are of particular interest.

\quad
This DELPHI analysis of the orbitally excited mesons \fo, \fd, \KOT\ and \fdp\ 
has recently been published~[1] and only some key points and final results 
will be discussed here. Tight selection criteria were required to achieve the 
highest possible purities of particle identification of the DELPHI Ring Imaging 
Cherenkov (RICH) detectors and Time Projection Chamber (TPC). Good agreement 
between the data and about 3 million detector simulated hadronic \ZO\ decays 
was observed.\\ 

\smallskip
{\bf 2. RESULTS AND DISCUSSION}

\smallskip
\smallskip
\smallskip

\begin{figure}[tbhp]
\centering\mbox{\epsfig{file=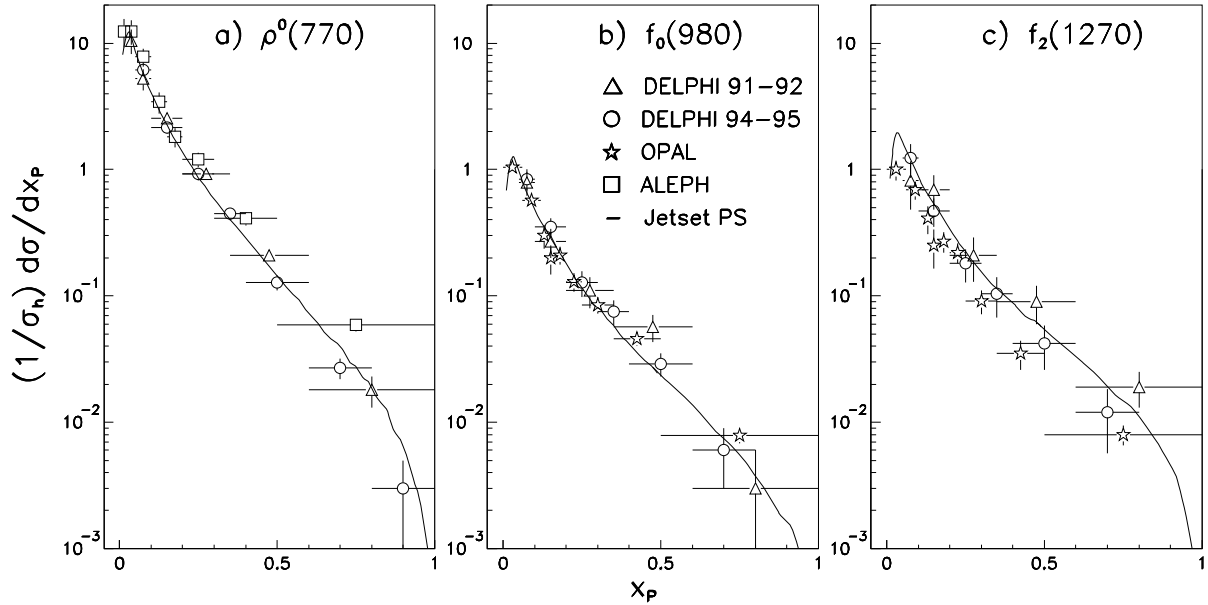,width=\linewidth}}
\caption[]{\small The differential cross-sections 
$(1/\sigma_h)(d\sigma/dx_p)$
for \ro, \fo\ and \fd\ mesons.}
\label{fig1}
\end{figure}

\quad
Figure~\ref{fig1} shows the final \xp\ distributions obtained for \ro, \fo\ 
and \fd, compared to a previous DELPHI analysis~[2] and to available data 
from ALEPH~[3], OPAL~[4] as well as the JETSET prediction. 
The previous DELPHI analysis was based on less statistics 
and did not apply particle identification.  
The DELPHI and OPAL results on the \fo\ \xp\ spectra agree quite well. 
The \fd\ \xp\ spectra measured by DELPHI and OPAL agree in 
shape but differ by 1.3 standard deviations in the normalisation. 
The agreement of the spectra in Figure~\ref{fig1} with the JETSET 
prediction is very satisfactory. The shapes of the \ro, \fo\ and 
\fd\ \xp\ spectra for $x_p \leq 0.4$ appear to be approximately the 
same. For $x_p > 0.4$, there is some indication that the \fo\ and 
especially the \fd\ \xp\ spectra are harder than the \ro\ \xp\ 
spectrum, in agreement with JETSET expectations.

\quad
The total production rates per hadronic \ZO\ decay for \ro, \fo, \fd, \KOT\
and \fdp\ are presented in Table~\ref{tab1}.
It is interesting to compare the total production rates 
of the tensor mesons with the respective rates of the vector mesons. 
Taking the \KOV\ and \ph\ total rates from~[6]:
$f_2(1270)/\rho^0 = 0.180 \pm 0.035$, 
${\rm K}^{*0}_2(1430)/{\rm K}^{*0}(892) = 0.095\pm 0.031$ and
$f^{\,'}_2(1525)/\phi = 0.115\pm 0.058$.
The differences between these ratios might indicate, that this is a simple 
consequence of the difference in particle masses and the mass dependence of 
the production rates. 
\begin{table}[tbhp]
\caption[]{\small The \ro, \fo, \fd, \KOT\ and \fdp\ production rates 
extrapolated to the full $x_p$-range (with the combined statistical and 
systematic errors) and their comparison with the predictions of the 
thermodynamical model [5].}
\begin{center}
\begin{tabular}{|l|c|c|}  
\hline 
Particle & DELPHI results & Thermodynamical model \\
\hline
\ro\  & $1.19  \pm 0.10 $ & $1.17   \pm 0.05  $ \\
\fo\  & $0.164 \pm 0.021$ & $0.0772 \pm 0.0076$ \\ 
\fd   & $0.214 \pm 0.038$ & $0.130  \pm 0.015 $ \\   
\KOT\ & $0.073 \pm 0.023$ & $0.0462 \pm 0.0041$ \\
\fdp\ & $0.012 \pm 0.006$ & $0.0107 \pm 0.0007$ \\ 
\hline
\end{tabular}
\end{center}
\label{tab1}
\end{table} 

\begin{figure}[tbhp]
\noindent
\begin{minipage}{0.48\linewidth}
\centering\epsfig{file=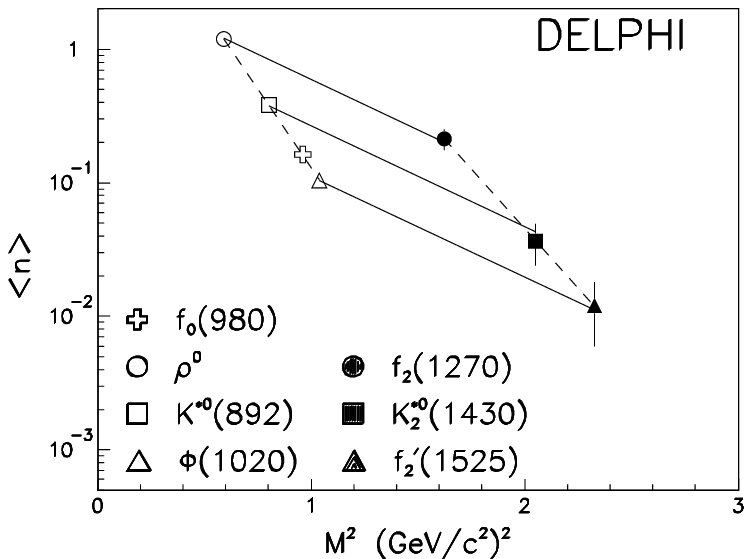,width=\linewidth}
\caption[]{\small The production rates of the scalar, vector and tensor 
mesons measured by DELPHI as a function of their mass squared.}
\label{fig2}
\end{minipage}\hfill
\begin{minipage}{0.48\linewidth}
\centering\epsfig{file=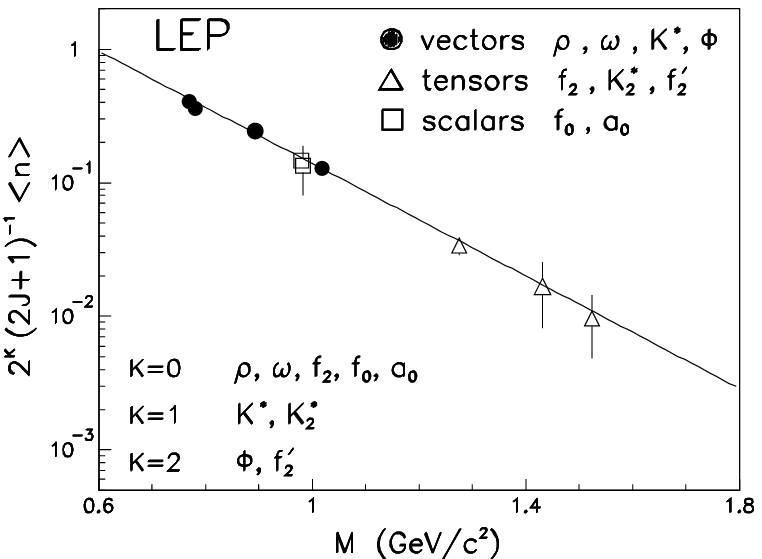,width=\linewidth}
\caption[]{\small The weighted production rates of the scalar, vector and 
tensor mesons measured by LEP as a function of their mass.}
\label{fig3}
\end{minipage}
\end{figure}

\quad
This suggestion is supported by Figure~\ref{fig2}, where the total rates 
measured by DELPHI for the \fo, \ro, \KOV, \ph, \fd, \KOT\ and \fdp\ 
are plotted as a function of their mass squared. 
Anti-particles are not included in the \KOV\ and \KOT\ rates. Both the 
scalar and vector data points and the tensor data points are well described by 
exponentials of the form $Ae^{-BM^2}$ (dashed lines). Figure~\ref{fig2} also 
shows that the mass dependence of the production rates is almost the same for 
three sets of data points: \ro\ and \fd ; \KOV\ and \KOT ; \ph\ and \fdp. 
These points are well fitted to the exponentials $Ae^{-BM^2}$ (solid lines), 
with three different normalisation parameters $A$ but the {\it same} slope 
parameter $B$.
Thus the relation between the production rates of tensor and vector 
mesons indeed appears to be very similar for different particles
if the mass dependence of these production rates is taken into account. 

\quad
The total production rates of the tensor mesons \fd, \KOT\ and 
\fdp\ are found to be rather small in absolute value, 
when compared with the vector meson production rates. This agrees
with common expectations that the production of 
orbitally excited states is suppressed. Ho\-wever, recently it was noticed 
[7] that the production rates of orbitally excited mesons 
are larger by factor of 5 than those of states with no orbital 
momentum if the universal mass dependence of the production rates 
for the pseudoscalar and vector mesons and the octet and decuplet 
baryons is accounted for. 

\quad
Another indication for the excess of orbitally excited mesons 
can be seen from Table~\ref{tab1}, where a comparison of the data with the 
recently proposed thermodynamical model [5] is presented. 
This model provides a very good description of the total production 
rates for the pseudoscalar and vector mesons and for the octet and decuplet 
baryons. This is illustrated in Table~\ref{tab1} by a very good
agreement between the model prediction and the data for the \ro. 
However, comparison of the model predictions with the present DELPHI 
results for the total production rates of orbitally excited mesons 
indicates that the model underestimates their yields by
about the same factor of 1.6--2.1, except for the \fdp, where the 
experimental uncertainties are quite large.

\quad
The large excess of orbitally excited mesons with respect to the vector
mesons can be explained by introducing the extra suppression factor $2^{-k}$, 
where $k$ is the number of $s$ and $\bar{s}$ quarks in the meson, as well as
the spin factor $2J+1$. This is seen from Figure~\ref{fig3}, where the total
production rates of the orbitally excited [1,4,8,9] and vector (see [7] and 
references therein) mesons averaged from all available LEP measurements and 
weighted with factor $2^{\,k}\,(2J+1)^{-1}$ are shown as a function of their 
mass. All data points are well described ($\chi^2/ndf$=4.8/7) by exponential 
of the form $2^{\,k}\,(2J+1)^{-1}{\langle n \rangle} = Ae^{-bM}$, with the 
parameters $A = 17.8 \pm 3.9$ and $b = 4.85 \pm 0.24$ (GeV/$c^2$)$^{-1}$.
Thus there is no excess of orbitally excited mesons with respect to the vector
mesons, if the dependence of their production rates on mass ($M$), spin ($J$)
and number of $s$ and $\bar{s}$ quarks ($k$) is accounted for.\\ 

\smallskip
{\bf 3. SUMMARY}

\smallskip
\smallskip
\smallskip

\quad
The DELPHI results [1] on inclusive production of the \ro, \fo, \fd, \KOT\
and \fdp\ in hadronic \ZO\ decays at LEP have been presented. They are
based on a data sample of about 2 million hadronic events, using the particle 
identification capabilities of the RICH and TPC detectors, and 
supersede the previous DELPHI results [2], with which they are consistent.
The following conclusions can be drawn.
\begin{itemize}
\item The total \fo\ and \fd\ production rates per hadronic \ZO\ decay 
      are $0.164 \pm 0.021$ and $0.214 \pm 0.038$ respectively.
      The \fo\ and \fd\ momentum spectra are well described by the tuned
      JETSET model. The shapes of the \fo\ and \fd\ momentum spectra are 
      similar to that for the \ro\ for $x_p \leq 0.4$. For higher \xp\ 
      values there is some indication that the ratios \fo/\ro\ and 
      especially \fd/\ro\ may increase with \xp, in agreement with JETSET 
      expectations. The total \fo\ and \fd\ rates and their momentum spectra 
      are consistent with the OPAL measurements [4].
\item The total \KOT\ and \fdp\ production rates per hadronic \ZO\ decay 
      amount to $0.073 \pm 0.023$ and $0.012 \pm 0.006$ and are about half
      the size of the rates predicted by the tuned JETSET model. The total
      \KOT\ rate is smaller by 1.8 standard deviations than the value
      $0.238 \pm 0.088$ measured by OPAL [8] for $x_E\leq 0.3$ and extrapolated
      by us to the full $x_E$ range.
\item The ratios \fd/\ro, \KOT/\KOV\ and \fdp/\ph\ are $0.180 \pm 0.035$, 
      $0.095 \pm 0.031$ and $0.115 \pm 0.058$ respectively. They appear to
      be somewhat different. However, the relationships between the
      production rates of the tensor and vector mesons for the \fd\ and \ro,
      \KOT\ and  \KOV, \fdp\ and \ph\ are found to be very similar when the
      mass dependence of the production rates is accounted for.
\end{itemize}

\bigskip
{\bf REFERENCES}

\smallskip
\smallskip
\smallskip

\small
\smallskip
[1]~~DELPHI Collab., P.~Abreu et al., {\it Phys. Lett.} B {\bf 449} (1999) 364.

[2]~~DELPHI Collab., P.~Abreu et al., {\it Z. Phys.} C {\bf 65} (1995) 587.

[3]~~ALEPH Collab., D.~Buskulic et al., {\it Z. Phys.} C {\bf 69} (1996) 379.

[4]~~OPAL Collab., K.~Ackerstaff et al., {\it E. Phys. J.} C {\bf 4} (1998) 19.

[5]~~F.~Becattini, {\it Z. Phys.} C {\bf 69} (1996) 485;\, private 
communication.

[6]~~DELPHI Collab., P.~Abreu et al., {\it Z. Phys.} C {\bf 73} (1996) 61.

[7]~~P.V.~Chliapnikov,~V.A.~Uvarov, {\it Phys. Lett.} B {\bf 423} (1998) 401;\,
{\it ibid} {\bf 345} (1995) 313.

[8]~~OPAL Collab., R.~Akers et al., {\it Z. Phys.} C {\bf 68} (1995) 1.

[9]~~OPAL Collab., K.~Ackerstaff et al., {\it E. Phys. J.} C {\bf 5} (1998) 411.
\end{document}